\documentclass[letterpaper,twocolumn,10pt]{article}
\usepackage{usenix-2020-09}

\usepackage{tikz}
\usepackage{amsmath}

\usepackage{filecontents}

\usepackage{microtype}
\usepackage{fancyvrb}
\usepackage{tcolorbox}
\usepackage{booktabs}
\usepackage{listings}
\usepackage{graphicx}
\usepackage{etoolbox}
\usepackage{xspace}
\usepackage{blindtext}
\usepackage{etaremune}
\usepackage{color}
\usepackage{algorithm}
\usepackage{algpseudocode}
\usepackage{algpseudocode}

\usepackage{caption}
\usepackage{subcaption}

\newcommand{\scheme}{RegGuard\xspace}
\newcommand{\codechunk}[1]{\textsf{\small #1}}

\begin{document}
%-------------------------------------------------------------------------------

%don't want date printed
\date{}

\title{\scheme: Leveraging CPU Registers for Mitigation of \\Control- and Data-Oriented Attacks}

%for single author (just remove % characters)
\author{
{\rm Munir Geden}\\
Department of Computer Science\\
University of Oxford, United Kingdom\\
{\tt munir.geden@cs.ox.ac.uk}
\and
{\rm Kasper Rasmussen}\\
Department of Computer Science\\
University of Oxford, United Kingdom\\
{\tt kasper.rasmussen@cs.ox.ac.uk}
} % end author

\maketitle

\begin{abstract}
CPU registers are small discrete storage units, used to hold temporary data and instructions within the CPU. Registers are not addressable in the same way memory is, which makes them immune from memory attacks and manipulation by other means. In this paper, we take advantage of this to provide a protection mechanism for critical program data; both active local variables and control objects on the stack. This protection effectively eliminates the threat of control- and data-oriented attacks, even by adversaries with full knowledge of the active stack.

Our solution \scheme, is a compiler register allocation strategy that utilises the available CPU registers to hold critical variables during execution. Unlike conventional allocations schemes, \scheme prioritises the security significance of a program variable over its expected performance gain. Our scheme can deal effectively with saved registers to the stack, i.e., when the compiler needs to free up registers to make room for the variables of a new function call. With \scheme, critical data objects anywhere on the entire stack are effectively protected from corruption, even by adversaries with arbitrary read and write access.

While our primary design focus is on security, performance is very important for a scheme to be adopted in practice. \scheme is still benefiting from the performance gain normally associated with register allocations, and the overhead is within a few percent of other unsecured register allocation schemes for most cases. We present detailed experiments that showcase the performance of \scheme using different benchmark programs and the C library on ARM64 platform.
\end{abstract}

\section{Introduction}
Despite many years of effort, memory bugs continue to be one of the root causes of software security problems, especially in applications developed using languages like C and C++, which are heavily used for real-time and systems programming. Since there are no built-in mechanisms in those languages that prevent people from placing critical program data right next to untrusted user or environment input, an attacker exploiting a bug in the program (e.g.,~buffer overflow) can overwrite control and data objects beyond the abstraction given in the source code.

Several schemes have been proposed to mitigate the possible impact of these bugs. The majority of these focus on \textit{control-oriented} attacks in which code pointers are targeted. For example, stack canaries~\cite{Cowan1998StackGuardOf} place random values on the stack to detect overflows onto return addresses. But these canaries fail to catch well-targeted corruptions, e.g.,~format string attacks, that can target specific addresses and leave the stack canary untouched. More powerful control-flow protections exist that do not make assumptions about how memory corruption happens. They include techniques like shadow stack~\cite{Abadi2005} which detects attacks on (shadowed) control data, or safe stack~\cite{Kuznetsov2014Code-pointerIntegrity} that prevents control data from being attacked. Such control-flow protections often do not address \textit{data-oriented} attacks that only target non-control data, for example, a condition variable controlling the execution of a privileged branch. Proposed data-flow protections against those attacks, e.g.,~data-flow integrity (DFI)~\cite{Miguel2006SecuringIntegrity}, check non-control data in addition to code pointers, and in the process introduce high performance costs.

Regardless of their limitations, current proposals for control- and data-oriented attacks face three common challenges in general. The first one is the performance overheads due to the instrumentation that accompanies legitimate memory operations. The second challenge is that their success is dependent on how well the instrumentation data (e.g.,~shadow stack) or segregated data (e.g.,~safe stack) is protected within the same program space. Current techniques hide the location of those through randomisation or implement some access policies for them. However, integrated attacks that reveal or search the location of instrumentation data can break the schemes' promises~\cite{Evans2015MissingIntegrity,Goktas2016BypassingProfit}. The final third issue is the lack of deployability by different device types and architectures. For strong assurance, many proposals either require instruction set (ISA) modifications~\cite{Davi2015,Christoulakis2016,Song2016HDFI:Isolation,Nyman2017} or require features provided by a specific architecture (e.g.,~Intel MPX~\cite{Burow2019SoK:Stacks}), which makes them deployable only for future devices or a small portion of existing systems. Also, the majority of defences are designed for high-end devices with a reliable operating system, whereas primitive architectures and systems (e.g.,~bare-metal) are generally ignored.

This paper presents \scheme, a novel scheme that leverages CPU registers to protect critical program data with additional assurance even if their states are saved to the stack. Our scheme successfully addresses all three challenges mentioned above and differs from previous proposals by providing practical and robust protection against both control- and data-oriented attacks. It is practical because \scheme is designed as an instrumentation only scheme that does not require any new hardware. It is robust because CPU registers, as unaddressable storage units, provide a strong hardware root of trust for the storage of critical data. Thanks to our cryptographic integrity assurance on saved register states, \scheme does not need to worry about integrated attack scenarios as it does not generate any instrumentation data that must be hidden or protected in program memory. 
Lastly, because \scheme is built on one of the fundamental building blocks of computers (i.e.,~CPU registers), it can be adapted to different device types and architectures, including both modern and legacy systems, with trivial changes on running software stack.

In order to verify that the integrity checks introduced by \scheme do not make the performance of the resulting binary unusable, we implement \scheme using LLVM compiler for the ARM64 architecture. ARM has been one of the most dominant architectures of mobile phones and microcomputers for a while, which makes it a good platform for testing performance. Our results show that for many applications compiled with \scheme, the performance is within a few percent of a normal optimised (\codechunk{-O2}) binary. 

\section{Background}
This section provides background information about the attack classes that \scheme mitigates, and explains how conventional register allocation schemes work.

\label{sec:background}
\subsection{Control- and Data-Oriented Attacks} 
Even if the integrity of the program code is assured and the stack is set as non-executable, the attacker can still perform on the memory in many ways. The first option is to hijack the program’s control flow for a code-reuse attack by targeting \textit{control data}. By carefully crafting code pointers, the attacker can express his attack from existing instructions and execute them with the order and data he would benefit from. To achieve this, he can modify return addresses, e.g.,~return-oriented programming (ROP)~\cite{Checkoway2010}, or indirect branch targets, e.g.,~jump-oriented programming (JOP)~\cite{Bletsch2011}, which we describe as \textit{control-oriented} attacks in general. Control-flow protections mitigate those scenarios by checking~\cite{Abadi2005} or ensuring~\cite{Kuznetsov2014Code-pointerIntegrity} the validity of control data. However, these techniques fall short of protecting against scenarios where the attacker corrupts only program variables without touching any code pointers. Such \textit{data-oriented} (non-control data) attacks~\cite{Chen2005Non-control-dataThreats} enable the adversary to reach his goal indirectly, for instance, by corrupting a condition variable that decides on a privileged branch execution (also called control-flow bending attacks~\cite{Carlini2015}). Apart from specific scenarios, those attacks can be Turing-complete with data-oriented programming (DOP)~\cite{Hu2016Data-OrientedAttacks,Ispoglou2018BlockAttacks} techniques in case of a suitable vulnerability, without disturbing control-flow protections. For a DOP scenario, the attacker must exploit a bug that can compromise a loop (the dispatcher) providing necessary branches and instructions (attack gadgets).

\subsection{Register Allocation}
Because accessing CPU registers is much faster than the memory, the compiler prefers mapping all program variables to the registers for better execution times. However, there is no practical constraint on the number of variables that can be defined in a program, despite the limited number of registers (i.e.,~usually no more than 32 general-purpose (GPR) and 32 floating-point registers (FPR) on modern architectures). Hence, a register allocation scheme must decide on how to share out registers to the variables. Thankfully, not all variables are concurrently live (i.e.,~code scope describing a variable definition to its final use) throughout the program execution. The compiler can thus utilise registers more efficiently by assigning the same registers to different variables (i.e.,~live ranges) at different times. 
If the number of live variables is more than available registers at any program point, called high \textit{register pressure}, the compiler handles those situations by \textit{spilling} some variables into the memory. The allocation scheme usually decides which variable to be spilled using \textit{spill costs} that estimate the candidate's number of uses and definitions, weighted proportionally to its loop nesting depth. The compiler can also store a variable both in the memory and registers by \textit{splitting} the live ranges for better utilisation.

\subsubsection{Allocation Level}
Register allocations can happen at a basic block, function or program level. If the basic block is chosen as the optimisation boundary, such an allocation scheme is called \textit{local} register allocation. Since local allocations~\cite{L.P.Horwitz1966} save and restore registers at basic block sites without taking into account the control-flow graph, they are not considered as optimal as \textit{global} allocations happening over the whole function. On the other hand, interprocedural (program-wide) allocations can only be meaningful for small programs with many short procedures. Therefore, global register allocations are generally used in practice. Global allocators enable reusing the same register file repeatedly for each function call. Depending on the calling convention in place, if a register is described as a \textit{caller-saved} register, its state is saved/restored at call sites by caller functions. Otherwise, the function to be called is responsible for saving and restoring a \textit{callee-saved} register. These operations are mostly performed through simple push-pop instructions as part of the callee’s prologue (save) and epilogue (restore) code.

\subsubsection{Global Allocation Techniques}
\label{sec:global-schemes}
Global schemes can adopt different approaches to solve the allocation problem. Graph colouring~\cite{Chaitin1981RegisterColoring,Chow1990,Briggs1994} is the most popular technique. It starts by building an interference graph, where the nodes represent variables and the edges connect two simultaneously live variables. The problem is formulated as two adjacent (interfering) nodes (variables) cannot be coloured with the same colour (register). Since the given problem is NP-complete, heuristic methods are used to solve the problem. For a graph, the degree of which is greater than the number of available colours (registers), meaning register pressure, the compiler can spill some candidates to the memory based on their \textit{spill costs}, which estimates the performance loss of mapping a variable to memory based on use densities. The compiler can also iteratively transform the graph (code) in different ways to find a solution. For instance, it can \textit{split} a live range of a variable which creates additional nodes that reduce the degree of a node. Or it can \textit{coalesce} (merge) some non-interfering nodes that represent variable-to-variable operations, the total degree of which must still be less than available colours (registers).

As an alternative to graph-colouring, linear scan~\cite{Poletto1999} techniques aim for faster compilation times. As the name implies, they generally maintain an active list of variables that are live at the current point in the function, the intervals of which are chronologically visited to perform register allocations. This allows linear scan techniques to handle interferences without computing a graph. Those techniques~\cite{Wimmer2010} can especially benefit from single static assignment (SSA) features that reduce the time spent in data-flow analysis and liveness analysis, thanks to unique variables defined on each assignment. Because naive techniques do not backtrack, they might result in less optimal allocations. However, proposals such as \textit{second-chance binpacking}~\cite{Traub1998QualityAllocation} address this by utilising lifetime holes (e.g.,~a scope where the value is not needed) of register values, which allows a spilled value to be placed on a register back again (splitting).

\section{Problem Setting}
\label{sec:problem}
On most systems, it is common to separate program memory into different segments and enforce different access patterns on each segment depending on what is stored there. For example, code (text) segments can be marked read-only and data segments can be marked non-executable without losing any functionality. This has made it harder to execute simple buffer overflow attacks and made more sophisticated control- and data-oriented attacks more prevalent. Although there exist control-flow protections~\cite{Abadi2005,Kuznetsov2014Code-pointerIntegrity} that mitigate control-oriented attacks, they fail to capture the more challenging instances of data-oriented attacks where the attacker only modifies non-control data, e.g.,~condition variables.

Addressing those scenarios has proven difficult to deploy in practice as they either introduce heavy instrumentation costs~\cite{Miguel2006SecuringIntegrity} or require expensive hardware changes~\cite{Song2016HDFI:Isolation}. Furthermore, software-based approaches against both attack classes have to ensure the integrity of their instrumentation data within the same memory space. Commonly used hiding mechanisms such as randomisation can be circumvented when the location of the data is revealed through an integrated attack. This paper addresses both control- and data-oriented attacks while taking into those drawbacks account.

\begin{figure}[tp]
\centering
\includegraphics[scale=0.21]{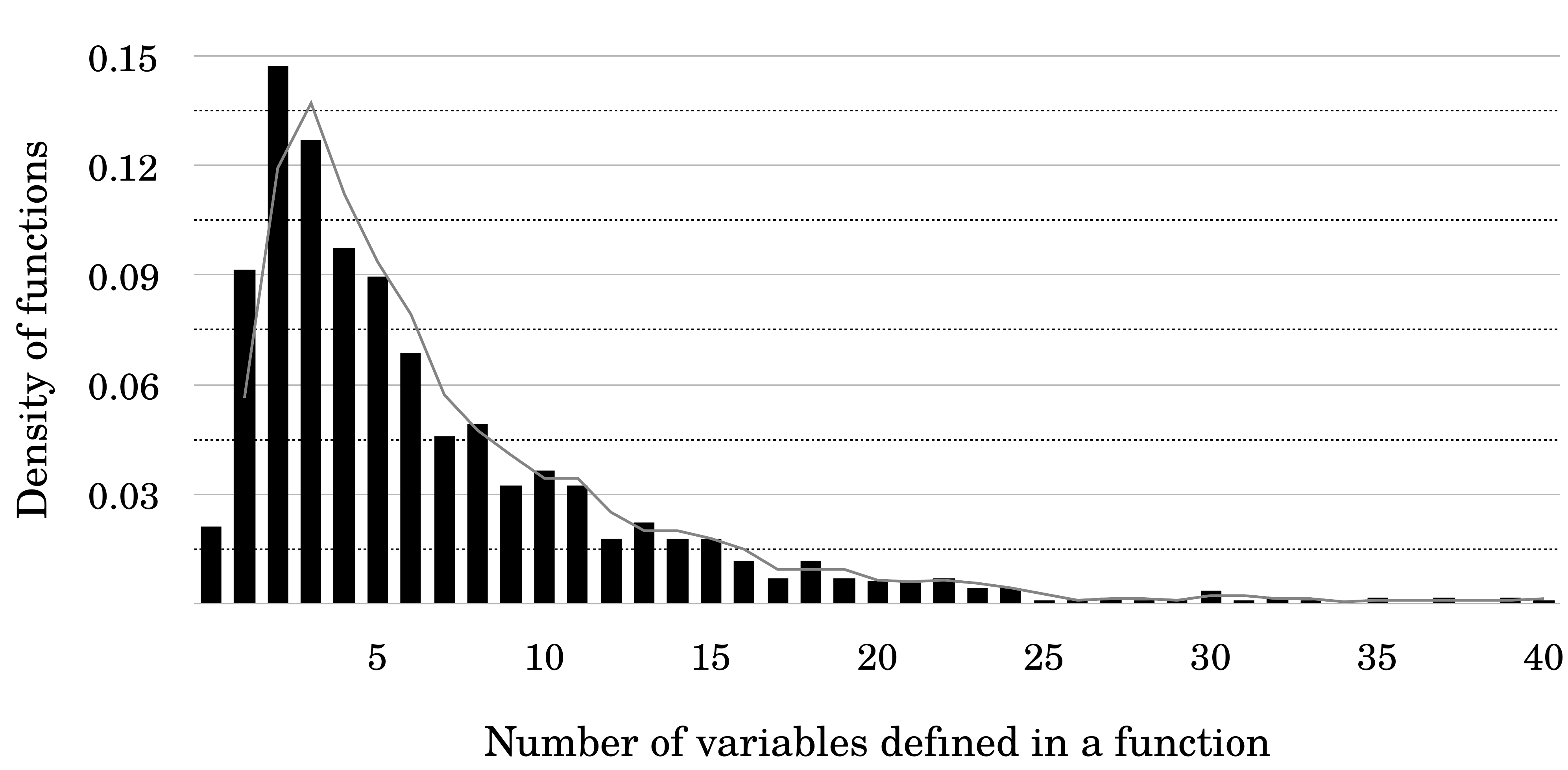}
\caption{Probability distributions of variable counts.}
\label{fig:distribution}
\end{figure} 

\subsection{Motivation}
For a successful control- or data-oriented attack, the attacker must either overflow some memory buffer onto the target object (i.e.,~relative address attack) or take over a data pointer for accessing the ultimate target (i.e.,~absolute address attack). CPU registers are immune from such attacks since they cannot be addressed in the same way.

However, to use CPU registers as a protection mechanism, we have to solve a couple of challenges. First, we have to find a way to use them for security while still allowing them to serve their primary purpose, namely as a fast storage mechanism for values in use to reduce execution time. Second, we have to find a way to leverage the limited capacity of the registers to protect all the relevant variables in a program. Simply using CPU registers as program-wide storage (interprocedural allocations) would put a hard limit on the number of variables that a program can use, which is not a practical constraint. At the same time, register states that are saved to the stack, e.g.,~during function calls, void their immunity against potential corruptions. Hence, we need a global allocation scheme that can employ the same registers for each call without undermining their security. With such an integrity assurance, CPU registers can provide enough storage to protect critical control and data objects of the entire program.

To provide insight into the coverage such a protection scheme can provide, Figure~\ref{fig:distribution} shows the number of variables per function in a representative set of benchmark programs. We use the same set of programs for our performance evaluation in Section~\ref{sec:evaluation}. As seen, 93\% of functions have less than 16 variables, and 99\% has less than 32 variables. Considering the average number of variables (6.9) and arguments (2.6) found per function, most modern CPUs provide enough registers (with 16/32 GPRs and 32 FPRs) to secure those objects as potential attack targets. Note also that the counts represent all reference and primitive variables found in a function at any point, and do not take the liveness of variables into account, so the actual numbers will be smaller on average. In Section~\ref{sec:allocations}, we show how it is still possible to deal with the rare event that the number exceeds the number of available registers.\looseness=-1

\begin{figure}[tp]
\centering
\includegraphics[scale=0.9]{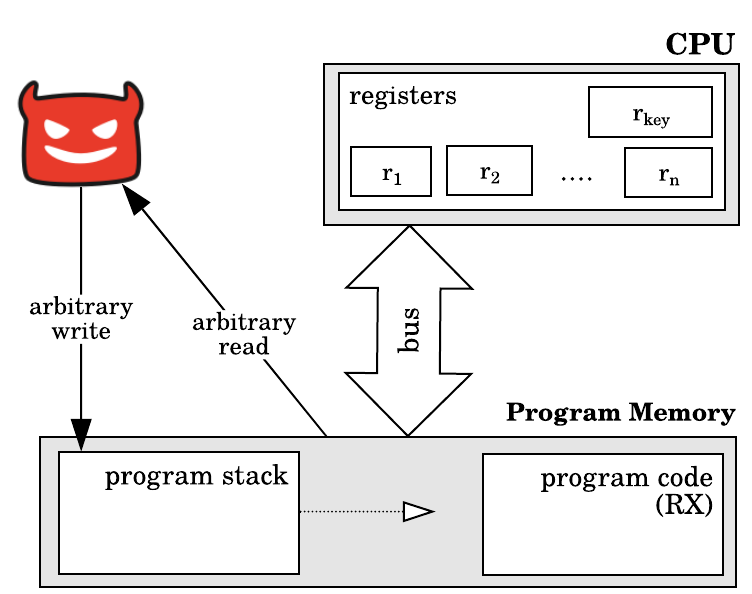}
\caption{Overview of the system components.}
\label{fig:system-model}
\end{figure} 

\subsection{System and Adversary Model}
In our model, the CPU is trusted and provides limited but secure register storage. Regarding the program memory, the system (see Figure~\ref{fig:system-model}) ensures code integrity via non-writable (RX) addresses, which can be provided by a flash memory or page-level write-xor-execute (W$\oplus$X) protections, depending on the device setting. 
The CPU has $n$ registers available (\textit{r\textsubscript{1}-r\textsubscript{n}}) for the scheme. 
The system dedicates a specific register (\textit{r\textsubscript{key}}) to store the key, for instance, a single FPR that is never saved to the program memory. We deliberately avoid making assumptions about the device type and its architecture. It can be a single-threaded bare-metal environment as well as a multi-threaded one with an operating system, the kernel space of which is trusted by the user processes. As long as the system has the necessary CPU registers and ensures the integrity of program code, our scheme is applicable to different architecture and software/firmware instances. We assume that the software stack running on the device can be recompiled and modified as required, without asking for any change in hardware.\looseness=-1

The adversary's goal is to manipulate the program runtime by modifying critical control and data objects on the stack, although program termination does not constitute a successful attack. For instance, he can target a code pointer such as a return address or a function pointer to hijack the program's control flow. Alternatively, he can overwrite a non-control data object, for example, a condition variable to manipulate the control flow indirectly. We assume a powerful adversary that has full read access to any part of memory at all times (including the stack), as well as write access to any address on the program stack. We are not going to explore how such read and write capabilities can be achieved in practice; we just grant the adversary this power. We do assume that the adversary cannot intervene in the compilation process and cannot modify program code in the non-writable code segment, which includes our instrumentation as part of it. 

This model captures both control- and data-oriented attacks extensively. It addresses both code-reuse attacks bypassing DEP, and more challenging data-oriented attacks that can otherwise circumvent control-flow protections. This model also covers a wide range of scenarios on how the adversary can interact with the program memory. In contrast to protections relying on random values (e.g.,~stack canary~\cite{Cowan1998StackGuardOf}) or random addresses (e.g.,~safe stack~\cite{Kuznetsov2014Code-pointerIntegrity}, ASLR~\cite{team2003pax}), this model covers integrated attacks~\cite{Goktas2016BypassingProfit} (e.g.,~memory disclosure) that can bypass those defences. 

\section{Design of \scheme}
\label{sec:design}
During the compilation process from source code down to machine code, the compiler has to map variable objects to either memory addresses or CPU registers. Since registers are safe from memory corruption and can be accessed very fast, we would prefer to put all variables in registers. However, this is not always possible as there can be more (simultaneously live) variables than available registers (i.e.,~high register pressure). Therefore, we must first ensure that the compiler prioritises a data object that is more likely to be targeted by the attacker for register allocation. Second, even if all variables of a function are assigned to registers, their values will be saved to the program stack during a function call, to make the registers available to the new function. Because these saved values can be overwritten on the stack, we must do something to guarantee their integrity.

\subsection{Security-Oriented Allocations}
\label{sec:allocations}
To ensure a register is primarily allocated to a variable that is more likely to be attacked, we assign a \textit{security score} to each variable. In contrast to conventional spill cost that estimates the performance burden of a variable left in the memory, a security score is a compile-time estimate of how critical a function variable is for the program runtime integrity. Variables with higher security scores are thus prioritised for register allocation and are included in the integrity checks designated for saved registers during a function call, explained in detail in Section~\ref{sec:protected-spills}.

\begin{figure}[tp]
\footnotesize
\centering
{
\lstset{
language=C,
basicstyle=\sffamily,
numbers=left,
frame=bt,
numberstyle=\footnotesize,
columns=fullflexible,
showstringspaces=false,
escapeinside={<@}{@>}
}      
\begin{lstlisting}
<@\dots@><@\textbf{\textit{high register pressure}}@><@\dots@>
int (<@\textcolor{blue}{*func\_ptr})@>(const char *,...) = &printf; /*function pointer*/ 
int <@\textcolor{blue}{is\_valid}@>=0;  /*decides on control flow*/
int drop_stats=0;  /*no critical use*/;
int max_trial=read();  /*user defined data*/
char data[SIZE];  /*buffer hosting untrusted environment data*/
/*the user has a legitimate control over the loop iterations*/
while (--max_trial>=0){
    /*vulnerable function*/
    read_buffer(data);
    if (check(data){
        <@\textcolor{blue}{is\_valid}@>=1;
        break;
    }
    drop_stats++;
}
if (<@\textcolor{blue}{is\_valid}@>==1)  /*decides on control flow*/
    do_process(data);  /*critical task*/
(<@\textcolor{blue}{*func\_ptr}@>)("trials of %s is %d", data,drop_stats);  /*print stats*/
<@\dots@><@\textbf{\textit{high register pressure}}@><@\dots@>
\end{lstlisting}
}
\caption{Code under register pressure for the given scope. For security, registers are allocated to \textcolor{blue}{\codechunk{func\_ptr}} and
\textcolor{blue}{\codechunk{is\_valid}} first instead of less critical \codechunk{max\_trial} and \codechunk{drop\_stats}.}\label{fig:raexample}
\end{figure}

\subsubsection{Security Scores}
\scheme considers variables listed below as primary attack targets that must be prioritised during register allocations. It assigns a security score to each according to the given order (i.e.,~the first in the list has a higher score).

\begin{enumerate}
    \item pointers, e.g.,~function pointers,
    \item programmer-defined values, e.g.,~\codechunk{is\_admin=1}
    \item condition variables, e.g.,~\codechunk{if(authenticated)}
\end{enumerate}

Pointers have the highest scores as they are the most common attack vector for powerful attacks. If not caught, the corruption of a code pointer such as an indirect branch or a call target can result in arbitrary execution, while a data pointer can be used to access or modify other data objects on the memory (i.e.,~absolute-address attack). Next comes the variables whose value is directly set by the programmer. These are prioritised over the variables whose defining agents are unknown function-wise because they represent the programmer's intentions as the legitimate program semantic. In contrast, the attacker would likely not benefit as much from corrupting a data object that is already controlled/defined by the user or environment~\cite{Geden2020TRUVIN:Integrity}. Then, condition variables used to make branch decisions are given registers, even if their value origins are unknown. Return addresses, return values, and function arguments are also assigned to registers. But they are excluded from this scoring and selection process because the calling convention in place already dedicates registers for them.\looseness=-1

Figure~\ref{fig:raexample} exemplifies how our \textit{security scores} differ from conventional spills costs. This code depicts a high register pressure for the given scope. Normally, a conventional scheme would allocate registers to \codechunk{drop\_stats} or \codechunk{max\_trial} variables first for better execution times as they will be accessed by each loop iteration. However, from the security point of view, \scheme considers that \textcolor{blue}{\codechunk{func\_ptr}} and \textcolor{blue}{\codechunk{is\_valid}} are more worthy of register allocation. Alteration of \codechunk{func\_ptr} as a code pointer can result in illegitimate execution of sensitive system functions, whereas modifying \codechunk{is\_valid} flag, which is both a programmer-defined and a condition variable, would manipulate branch decisions as a data-oriented attack. On the other hand, \codechunk{max\_trial} defined from an external source (e.g.,~the user) or \codechunk{drop\_stats} that does not affect control-flow of the function are not identified as critical.

Differently from spill costs given based on the use densities, security scores that represent the likelihood of a register candidate to be attacked are designed as a fast intraprocedural static approximation considering the type of a variable, its value agents and use purposes. Hence, a security score must be associated uniformly with multiple live ranges of a variable. In other words, the scores should not be localised according to different ranges of a variable. Algorithm~\ref{alg:scores} shows how those security scores are calculated to rank register candidates in an order that would maximise the security by taking into account those properties.

\begin{algorithm}[tp]
	\small
	\caption{\small Pseudocode of security score calculations}
	\label{alg:scores}
	\begin{algorithmic}
		\Function{SecurityScore}{$var$}
		\State $var.{score} \gets 1$
		\If{$var.type$ is a pointer type} 
		\State $var.{score} \gets var.{score}+4$
		\If{$var.uselist$ has a branch instance} 
		\State $var.{score} \gets var.{score}+1$
		\EndIf
		\ElsIf {$var.type$ is an integral type} 
		\If{$var.deflist$ has an immediate assignment} 
		\State $var.{score} \gets var.{score}+2$
		\EndIf
		\If{$var.uselist$ has a comparison instance} 
		\State $var.{score} \gets var.{score}+1$
		\EndIf
		\EndIf
		\EndFunction
	\end{algorithmic}
\end{algorithm}

\subsubsection{Allocation Process}
As a global allocation scheme, \scheme works at the function level to reuse the same register file repeatedly for each call and accommodate more critical data objects on registers. The allocation technique to be used (see Section~\ref{sec:global-schemes} for different options) should be chosen based on the features of the compiler. For instance, the compilers using single static assignment (SSA) form, such as LLVM, generally implement custom linear techniques with faster compilation times, whereas other compilers can provide graph-colouring as the default option (e.g.,~GCC). We highlight that the choice of allocation method, where some compilers provide as a configurable option, is a separate issue from the problem \scheme addresses. And it does not have any impact on the applicability of our scheme as long as conventional spill costs are replaced by the security scores proposed. Any global allocation technique provided by the compiler can thus be preferred.\looseness=-1

We remind that registers are actually allocated to the live ranges of variables. A \textit{live range} describes the instruction or basic blocks scope ranging from a value definition to its all uses for the same definition. Live range definitions allow us to reuse a register for different variables whose ranges do not interfere with each other. A variable can have multiple live ranges with potential gaps in between, where each one starts with a new definition. The variable does not have to occupy a register during these gaps. Hence, the allocation schemes generally utilise those for more optimal allocations. Such cases also benefit our scheme without undermining its security promises since the attacker cannot benefit from overwriting a variable value that will be later redefined before its use. The attack surface thus gets smaller as the registers are utilised better. This can be especially meaningful for architectures suffering from register scarcity. 

\begin{figure}[tp]
\centering
\includegraphics[scale=0.55]{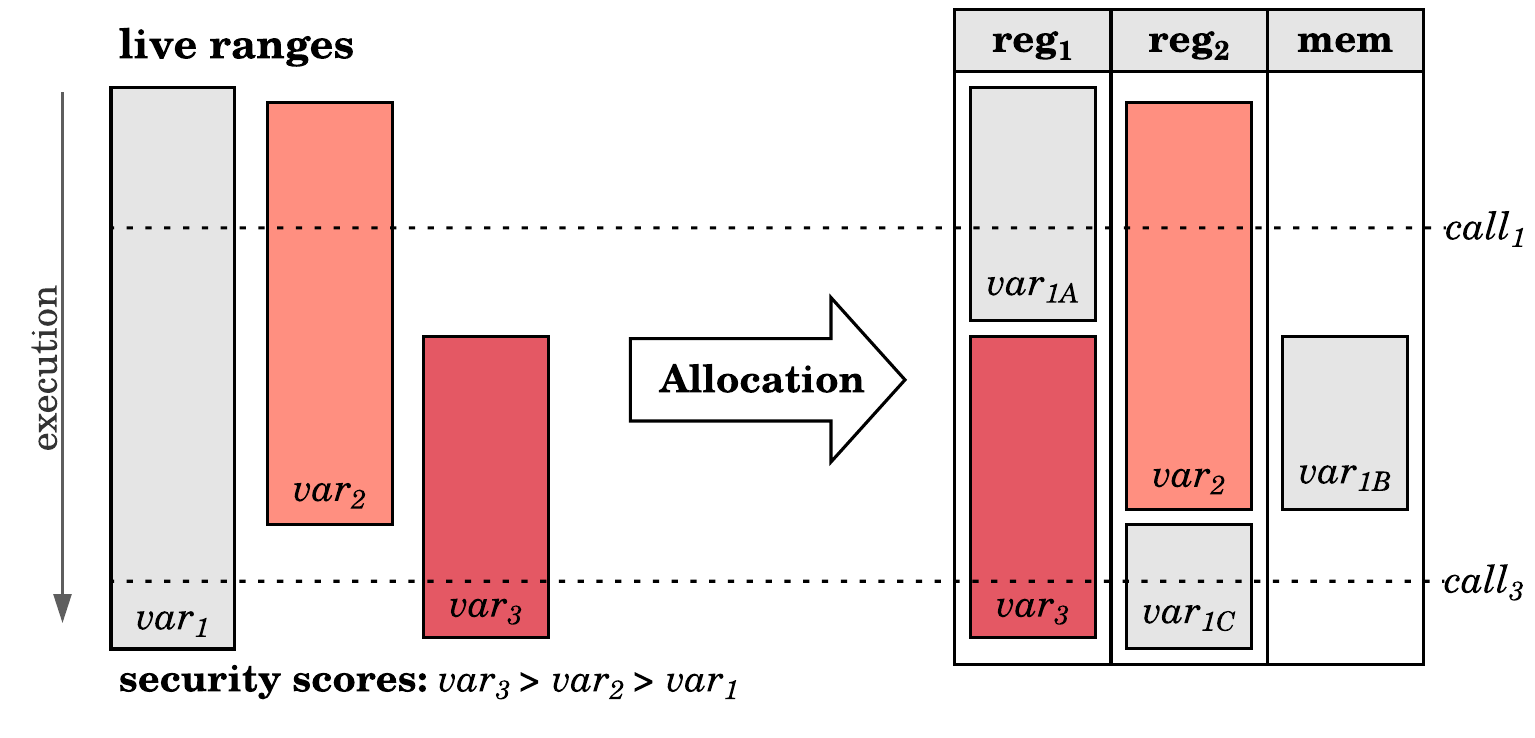}
\small
\caption{Allocations under register pressure.}
\label{fig:example}
\end{figure}

Figure~\ref{fig:example} depicts how \scheme should allocate available registers to the variables using security scores; so decides which variables to be protected. This example considers a scope under high register pressure with two available registers \textit{reg\textsubscript{1}} and \textit{reg\textsubscript{2}}, and three variables, the live ranges of which interfere as shown. The security scores are represented by colour tones, \textit{var\textsubscript{3}} is the most critical target, followed by \textit{var\textsubscript{2}}, whereas \textit{var\textsubscript{1}} has the lowest score. Using security scores, the scheme priorities two registers to \textit{var\textsubscript{3}} and \textit{var\textsubscript{2}} and spills \textit{var\textsubscript{1}} when required. However, the allocation method can still utilise gaps (i.e.,~instructions that \textit{var\textsubscript{3}} and \textit{var\textsubscript{2}} do not interfere), where a register become temporarily available for \textit{var\textsubscript{1}}. Those splits not only enhance the performance but also provide a better reduction of the attack surface. For instance, regardless of its criticality, \textit{var\textsubscript{1}} in the example can thus enjoy both performance and security promises, even if for a short time. And it will be safe during the execution of function calls, \textit{call\textsubscript{1}}, \textit{call\textsubscript{2}} depicted as potential attack vectors. Suppose such a case occurs while a critical variable range is left in the memory. In that case, the compiler would display a warning message to guide the programmer to review the code.\looseness=-1

\begin{table}[tp]
	\centering
	\small
	\begin{tabular}{@{}lr@{}}
		\toprule
		\textbf{Target Type}         & \textbf{Variance} \\
		\midrule
		Variables (Not Addressed) & Static      \\
		Variables (Locally Addressed) & Static     \\
		Variables (Called by Reference) & Dynamic      \\
		Temporaries & Static      \\ 
		Arguments & Static      \\ 
		Return Addresses & Static      \\ 
		Frame Pointers & Static     \\ 
		Return Values & Static    \\ 
		\bottomrule
	\end{tabular}
	\caption{Variance of register saves during the callee function.}
	\label{tab:stack-objects}
	
\end{table}

\subsection{Integrity of Saved Register Values}
\label{sec:protected-spills}

The program can save a register value to the stack for one of two reasons. The first is to free up a register for a more critical variable within the same function. These register spills can happen only under high register pressure, and the decision of evicting a register in use for another variable is guided by the security scores described in Section~\ref{sec:allocations}. The second more common reason, which we need to take care of, is a new function call that triggers the eviction of registers for the callee function. Those register states that belong to the caller's execution are saved to the stack either by the caller at call sites or by the callee as part of its prologue code. The decision of which registers must be saved/restored by the caller and the callee is mainly described by the calling convention. Regardless of the calling convention in use, any register state saved to the stack becomes vulnerable to memory corruption. Therefore, \scheme implements integrity checks on those to ensure that they are restored back to the registers without any corruption on return.

\subsubsection{Invariance of Saved States}
The integrity assurance covers saved register states that must not change during the execution of a callee function. Table~\ref{tab:stack-objects} presents an overview of those as potential targets. The only exception is the values that can be legitimately modified by the callee, usually an updateable value passed as a call by reference argument. Otherwise, \scheme protects local variables, return addresses, frame pointers, temporaries, function arguments, and return values that can be leveraged for an attack. With a fine-grained (e.g.,~flow-sensitive) pointer analysis~\cite{Hardekopf2011,Kuderski2019Unification-basedOversharing} that distinguishes local pointers from call by reference arguments, where the latter must be destroyed following the call instruction, \scheme can also ensure the integrity of locally addressed variables whose values must not change during the callee's execution. Because pointer analysis is a separate research problem that is orthogonal to our study, we will not discuss this issue further.

\begin{figure}[tp]
	\centering
	\includegraphics[scale=0.68]{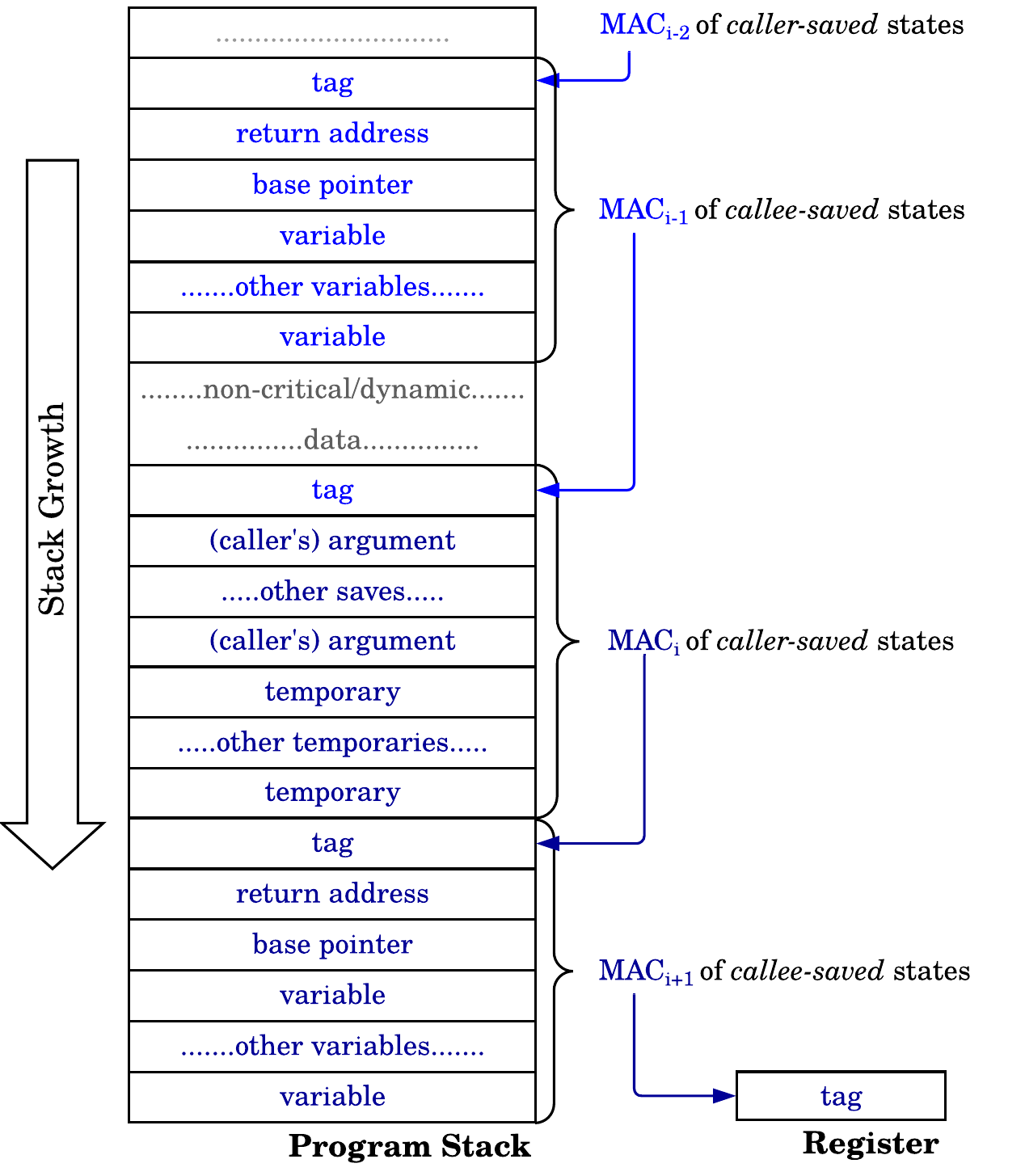}
	\caption{Securing saved register objects using a keyed hash.}
	\label{fig:secure-stack}
\end{figure} 

\subsubsection{Integrity Checks}
We recall that the data actively used on registers are already safe from attacks. Hence, the only attack surface left is register values saved to the stack. \scheme employs a cryptographic keyed hash (MAC) to guarantee that those saved register values have not been modified while they were on the stack. Prior to the execution of a function body, our scheme computes a reference tag from register objects being saved to the stack. This tag value is also kept on a specific register unless the callee function makes another call. Upon completion of the function body, a new tag is generated from actual objects being restored to the registers. This tag is compared against the reference tag previously generated from saved objects, any corruptions on those can thus be revealed. For a function call consisting of both caller- and callee-saved registers, this is a two-step process connected. The first tag generation/verification of caller-saved registers is managed at call sites, while the second tag digesting callee-saved registers is created/checked by the function prologues/epilogues. Function-wise, \scheme digests each call frame using a single tag value. Program-wide, because we save the tag register to the stack with other registers and include it in the next tag calculation, our scheme actually creates a chain of tags that provides (almost) a complete stack image on a single register. This prevents the attacker from replaying a (standalone) call frame and its corresponding tag for a different call context. We remind that the MAC key used is never saved to the same program/process memory, which is adequate to authenticate any tag restored from the stack that serves as the integrity proof of restored objects. Thanks to a single key kept on a register and MAC calculations that are also part of non-writable program code, \scheme provides an integrity-guaranteed storage mechanism for each call.

Figure~\ref{fig:secure-stack} depicts an overview of a call stack tied with tags. \scheme creates a tag for each callee- and caller-saved region, where the tag of a caller-saved region also contains the previous tag of a callee-saved region or vice versa. This helps us to bind frames to each other for a tight representation of the whole program stack. Equations~\eqref{eq:cksum-caller} and~\eqref{eq:cksum-callee} below express what each tag created for caller- and callee-saved regions contains.
{\small
\begin{align} 
\mathrm{tag}_{i}&= {\mathrm{MAC_{sk}}}(\mathrm{tag}_\textit{i-1}\mathbin\Vert\mathrm{arg}\textit{1}_{i-1}\mathbin\Vert...\mathbin\Vert\mathrm{arg}\textit{n}_{i-1}\mathbin\Vert...\mathbin\Vert\mathrm{tmp}\textit{n}_{i-1})\label{eq:cksum-caller}\\
\mathrm{tag}_{i+1}&= {\mathrm{MAC_{sk}}}(\mathrm{tag}_{i}\mathbin\Vert\mathrm{ret}_{i}\mathbin\Vert\mathrm{bp}_{i}\mathbin\Vert\mathrm{var}\textit{1}_{i}\mathbin\Vert ...\mathbin\Vert\mathrm{var}{n}_{i})\label{eq:cksum-callee}
\end{align}}%
Although the details can vary depending on the calling convention and the architecture, we consider the caller is responsible for saving and restoring its arguments (\textit{arg}) and temporaries (\textit{tmp}) at call sites while its return address (\textit{ret}), base/frame pointer (\textit{bp}) and local variables (\textit{var}) on registers saved by the callee. Even if the architecture (e.g.,~x86) does not use a link (return) register and stores the return address directly on the stack, it is still included in the tag generated for callee-saved regions as an object that must not be used until the return.

To reveal the corruption of a saved object, \scheme injects two groups of instructions. The first group generates a reference tag for register values being saved at function prologues and call sites. The second group checks whether this reference tag matches the one calculated from restored values. Both tag calculations directly align with existing register operations to avoid additional memory accesses. With a few scratch registers, \scheme can compute tags from directly register values. In order to make this possible, the compiler rearranges register restores in the same order they are pushed, instead of pop instructions working in the reverse order.

\subsubsection{Bootstrapping and Program Startup}
Regarding the bootstrapping of the system, the tag generation starts with the first call made by the software in question. For a simple setting with no process or privilege separation, such as a bare-metal or a RTOS environment, a single key to be shared by all tasks is generated at boot time using software or hardware RNGs available on the system. This key is assigned to an FPR dedicated as the key register. We note that this register is not saved to the memory by the scheduler or interrupt handler, thanks to the control over the software stack. If there is a hardware context switching in use, those instances also usually do not save FPRs. Otherwise, in the case of a general-purpose OS, a fresh key is generated at each process start. Only the kernel can save the key register to its own memory space, which is trusted by the user processes. User-managed threads share the same key and do not save the key register during a context switch.

\DefineVerbatimEnvironment%
{MyVerbatim}{Verbatim}
{fontsize=\footnotesize,fontfamily=helvetica,commandchars=\\\{\},
codes={\catcode`\$=3\catcode`^=7\catcode`_=8},frame=single}

\begin{figure}[tp]
\begin{MyVerbatim}
\textbf{read\_buffer:}                              \textit{#callee function}
    \textit{........................\textbf{prologue: register saves}.............................................}
    \textcolor{red}{INIT(key)}                                \textit{#initialise states (v\textsubscript{1-4}) with r\textsubscript{key}}
    store   r\textsubscript{tag},    [sp]
    \textcolor{red}{COMPRESS(m\textsubscript{1})}                   \textit{#m\textsubscript{1}=r\textsubscript{tag}}
    store   r\textsubscript{ret},    [sp-8]
    \textcolor{red}{COMPRESS(m\textsubscript{1},m\textsubscript{2})}             \textit{#m\textsubscript{2}=r\textsubscript{ret}}          
    store   r\textsubscript{bp},    [sp-16]
    \textcolor{red}{COMPRESS(m\textsubscript{2},m\textsubscript{3})}             \textit{#m\textsubscript{3}=r\textsubscript{bp}}
    store   \textcolor{blue}{r\textsubscript{var1}},  [sp-24]    
    \textcolor{red}{COMPRESS(m\textsubscript{3},m\textsubscript{4})}             \textit{#m\textsubscript{4}=r\textsubscript{var1}}
    store   \textcolor{blue}{r\textsubscript{var2}},  [sp-32]
    \textcolor{red}{COMPRESS(m\textsubscript{4},m\textsubscript{5})}             \textit{#m\textsubscript{5}=r\textsubscript{var2}}
    \textcolor{red}{r\textsubscript{tag} = FINALIZE(m\textsubscript{5})}
    sub     sp,    40
    \textit{........................body instructions...........................................................
    }
    ..............................................................................................................
   \textit{ ........................\textbf{epilogue: register restores}.........................................}
    mov     r\textsubscript{tmp1}, r\textsubscript{tag}                    \textit{#copy reference tag to a scratch register}
    \textcolor{red}{INIT(key)}                                \textit{#initialise states (v\textsubscript{1-4}) with r\textsubscript{key}}
    load    r\textsubscript{tag},    [sp+32]
    \textcolor{red}{COMPRESS(m\textsubscript{1})}                   \textit{#m\textsubscript{1}=r\textsubscript{tag}}
    load    r\textsubscript{ret},    [sp+24]
    \textcolor{red}{COMPRESS(m\textsubscript{1},m\textsubscript{2})}             \textit{#m\textsubscript{2}=r\textsubscript{ret}}          
    load    r\textsubscript{bp},    [sp+16]
    \textcolor{red}{COMPRESS(m\textsubscript{2},m\textsubscript{3})}             \textit{#m\textsubscript{3}=r\textsubscript{bp}}
    load    \textcolor{blue}{r\textsubscript{var1}},  [sp+8]    
    \textcolor{red}{COMPRESS(m\textsubscript{3},m\textsubscript{4})}             \textit{#m\textsubscript{4}=r\textsubscript{var1}}
    load    \textcolor{blue}{r\textsubscript{var2}},  [sp]
    \textcolor{red}{COMPRESS(m\textsubscript{4},m\textsubscript{5})}             \textit{#m\textsubscript{5}=r\textsubscript{var2}}
    \textcolor{red}{r\textsubscript{tmp2} = FINALIZE(m\textsubscript{6})}
    \textcolor{red}{CHECK(r\textsubscript{tmp1},r\textsubscript{tmp2})}               \textit{#check whether the checksums match}
    add     sp,    40
    ret
\textbf{example():}                                \textit{#code in Figure 2}
    \textit{........................instructions....................................................................
    }
    mov    \textcolor{blue}{ r\textsubscript{var1}}, &printf               \textit{#int (*func\_ptr)...; line 2}
    mov    \textcolor{blue}{ r\textsubscript{var2}}, 0                       \textit{#int is\_valid=0; line 3}
    \textit{........................instructions....................................................................
    }
    call      read\_buffer
    \textit{........................instructions....................................................................
    }
    mov     \textcolor{blue}{r\textsubscript{var2}}, 1                       \textit{#is\_valid=1; line 12}
    \textit{........................instructions....................................................................
    }
    cmp     \textcolor{blue}{r\textsubscript{var2}}, 1                       \textit{#if (is\_valid==1) line 17}
    \textit{........................instructions....................................................................
    }
    call       \textcolor{blue}{r\textsubscript{var1}}                           \textit{#(*func\_ptr)(...); line 19}
    \textit{........................instructions....................................................................
    }
 \end{MyVerbatim}
 \caption{MAC \textcolor{red}{calculations} aligned with register operations for the slice of \textcolor{blue}{\codechunk{func\_ptr}} and \textcolor{blue}{\codechunk{is\_valid}} variables.}
 \label{fig:manuel-example}
 \end{figure}

\subsubsection{Choice of MAC} 
The MAC function to be used should be chosen based on available features of the CPU architecture. If the ISA provides relevant vector and cryptographic extensions, we recommend using HMAC-SHA256 with hardware acceleration. Otherwise, we suggest using SipHash~\cite{Aumasson2012SipHash:PRF} as an architecture-agnostic option for CPUs that lack of cryptographic instructions. SipHash is a keyed hash primarily designed to be fast, even for short inputs, with a performance that can compete with non-cryptographic functions used by hash tables. Thanks to its performance benefits, SipHash is highly practical and deployable on different architectures.

Figure~\ref{fig:manuel-example} sketches how \scheme aligns its MAC calculations with register operations at function prologues and epilogues using SipHash. Both sections start by initialising internal states (on scratch registers) generated from the key and constants. Next, it applies compression rounds on those states by XORing them with message blocks (values) already on registers. Lastly, it concludes tag generation with the final message block (register). The reference tag representing saved values is not pushed to the stack unless the function body calls another function. Prior to the epilogue, this reference value is moved to a scratch register; the epilogue can thus restore the previous tag to the dedicated register as a part of the restoring process. The reference tag moved to a temporary register will be later compared against the actual tag generated from restored registers at the end before return. Any unmatch of two tags implies an attack because saved register objects cannot be changed unless the control is returned back to the caller function.\looseness=-1

\subsection{Security Analysis}	
As previously described, the adversary’s goal is to manipulate the program runtime by corrupting control and data objects on the stack. For the corruption to stay undetected, the adversary has to either skip the integrity checks or make those checks pass. We will look at each of these options in turn.\looseness=-1

In order to skip checks, the adversary must modify the binary to void the instrumentation. This is not possible in our model because the code segment is non-writable. For the adversary to pass integrity checks, he has to forge a valid tag or reuse a previously recorded one. Forging a valid keyed hash for an attack state either requires finding the second preimage of the legitimate state or access to the key. Since the key is protected on a register that is not saved to program memory, and therefore unavailable to the attacker, if the MAC-function is secure (i.e.,~existentially unforgable, and second preimage resistant), forging a valid tag without the key is only possible with negligible probability.

The adversary might attempt to replay a legitimate tag for a different call of the same or a different function. However, even with the same variable and argument values, replaying will not work. This is because each tag containing return address, base pointer and more importantly former tags (representing previous call frames) provides a very tight representation of the whole program stack, where the (most recent) tag digesting all context is kept safe on a register for a regular call stack with last in, first out (LIFO) order. Besides, replaying a tag for a different process execution is not an option since a fresh key is generated at each program start.

\begin{table}[tp]
	\centering
	\small
	\begin{tabular}{@{}lrr@{}}
		\toprule
		\textbf{Register}         & \textbf{Type}  & \textbf{Purpose}\\
		\midrule
		x0-x7 & Caller-saved   & Arguments   \\ 
		x9-x15 & Caller/e-saved   &  Temporaries \\ 
		x19 & Callee-saved   & Tag   \\ 
		x20-28 & Callee-saved   & Local variables   \\ 
		x29 & Callee-saved   &  Frame/base pointers  \\ 
		x30 & Callee-saved   &  Link/return addresses  \\
		q31 (FPR) & Reserved/not saved &  MAC key \\
		\bottomrule
	\end{tabular}
	\caption{The details of calling convention used.}
	\label{tab:cc-details}
\end{table}

\section{Implementation on ARM64}
\label{sec:implementation}
We have implemented a proof-of-concept\footnote{https://github.com/msgeden/llvm-project} of \scheme on ARM64 (AArch64) to evaluate its performance impact. \scheme can be adapted to different architectures such as x86, SPARC, MIPS, PowerPC and RISC-V (preferably 64-bit versions). But we have chosen ARM64 for demonstration purposes due to the following reasons:

ARM has been the dominant architecture of the mobile and embedded landscape for a long time. Thanks to Apple’s recently started transition to ARM-based processors and the embrace of Microsoft Windows, it is now projected that ARM will surpass Intel in the PC market in less than a decade~\cite{DavidFloyer2020ExitingPC}. Apart from promising market share, ARM64, with 31 GPRs (64-bit) and 32 FPRs (128-bit), has more registers than x64 (i.e.,~16 GPRs and 16/32 FPRs). Hence, even without having to modify the standard calling convention (ABI) of underlying software components, ARM64 provides enough registers to secure more variables than expected to be found per function (see Figure~\ref{fig:distribution}). For instance, the standard ABI dedicates 10 callee-saved registers compared to 6.9 variables found on average. Furthermore, registers reserved for arguments and temporaries not only help to secure other potential targets but also avoid register pressure in general. It also enables to use a FPR as two GPRs via vector form indexes. Besides, the ISA equipped with cryptographic extensions allows us to evaluate the hardware-accelerated HMAC-SHA256 option.

For the implementation, which consists of two parts, we have used the LLVM compiler, which is configured to dedicate a single FPR (128-bit) for the key and a GPR (64-bit) for tag values. For the first part, we have modified the \textit{basic} register allocation pass provided as a custom technique using priority queues to eliminate strict visits in linear order. Since we have not encountered register pressure on benchmark programs, thanks to plenty of registers available, our allocation pass simply ensures that registers given to variables are not spilled for performance reasons. For the second phase, we have mainly worked on the part responsible for target-specific prologue and epilogue code. For the proof-of-concept, integrity checks are placed for only callee-saved registers that are primarily assigned to local variables by the allocator. But the registers known as caller-saved can also be included in tag calculations using the same instrumentation, thanks to the compilation flags available (e.g.,~\codechunk{-fcall-saved-x9}). Table~\ref{tab:cc-details} summarises the highlights of the calling convention used during our experiments.

For simplicity, we have encapsulated MAC calculations with two functions added to the C library in use\footnote{https://github.com/msgeden/musl}. The first one (\codechunk{\_\_register\_mac}) is injected to the end of the prologue and generates a reference tag from saved register values. The second one (\codechunk{\_\_register\_check}), which is placed at the beginning of the epilogue, creates another tag from the values to be restored and compares it against the reference value. In the case of unmatched values, which means an attack, it terminates the program. Both wrapper functions take the start address and the size of the region where registers are pushed as their arguments. The latter function additionally requires the reference tag for comparison. The instrumentation also handles the preservation of original arguments required by the actual callee function and the return values upon its completion at call sites of the wrapper functions. For optimisation purposes, we have avoided injecting these two functions to the leaf functions of the program as their frames cannot be modified in practice without another function call.

Differently from the ideal design proposed in Section~\ref{sec:protected-spills}, those wrapper functions use stack values instead of directly using register values. We remind that as a proof-of-concept implementation avoiding the complexity, the wrapper functions introduce additional cache hits. Hence, our performance discussion should be seen as an over-approximation, whereas a production-ready implementation based on the proposed design would have less performance overhead.

For MAC, we have implemented two options. The first option is HMAC-SHA256, backed by hardware acceleration. The second one is SipHash-2-4 producing a 64-bit tag, as a fast, practical, and deployable option for different architectures lacking advanced vector and cryptographic extensions. 

\begin{figure}[tp]
	\centering
	\includegraphics[scale=0.21]{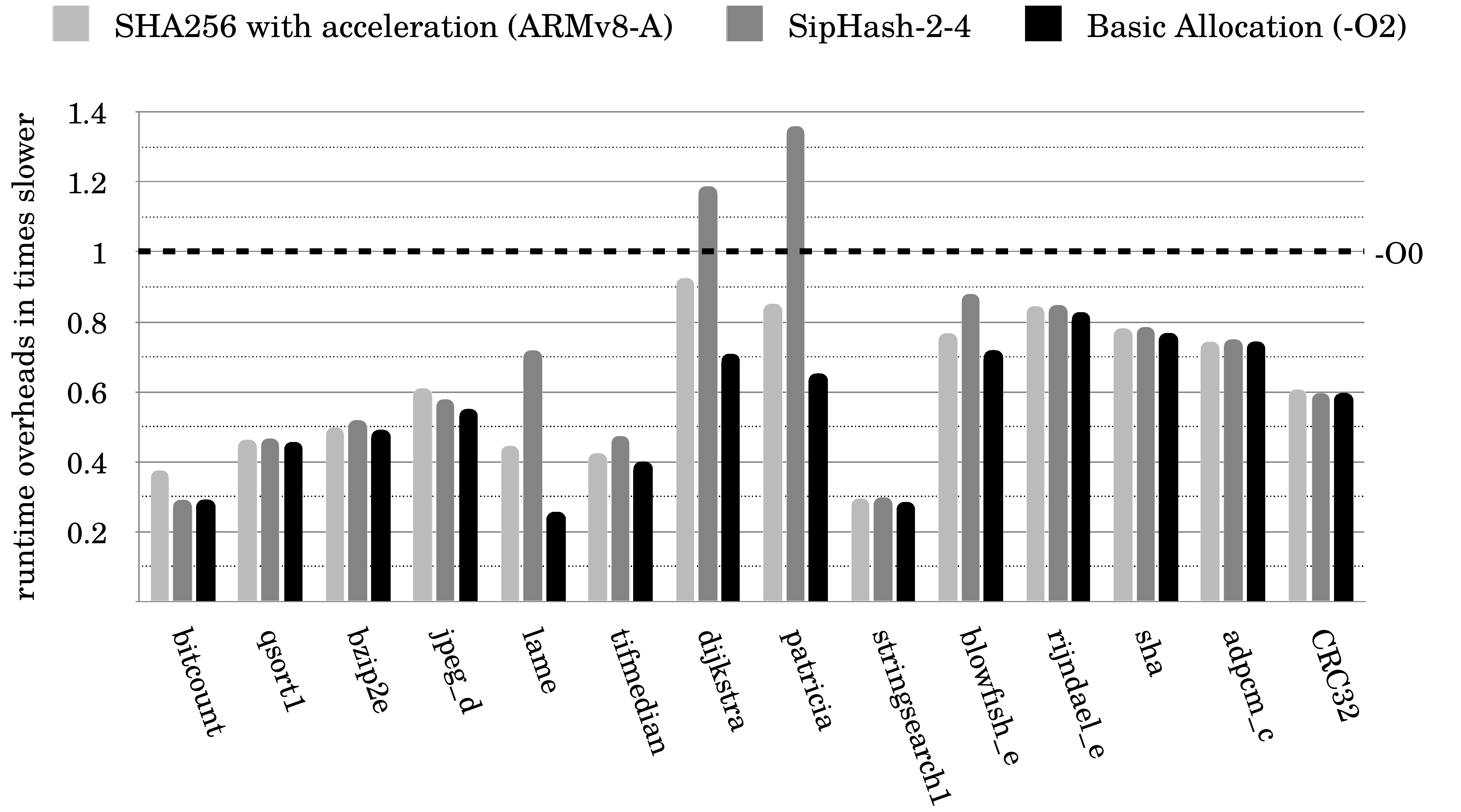}
	\caption{Runtime overheads of \scheme.}
	\label{fig:overheads}
\end{figure} 

\section{Performance Evaluation}
\label{sec:evaluation}
For performance evaluation, we have used \textit{cBench} suite that is derived from the popular open-source MiBench~\cite{Guthaus2001MiBench:Suite} benchmark suite. The experiments were performed with 14 C programs from different categories. We have run those programs on a Linux-AArch64 system running on an Apple device with an M1 chip. Instead of \textit{glibc} provided by the system, we have linked our benchmark executables to \textit{musl libc} as the C library. We have instrumented not only benchmark executables but also the C library interacting with the kernel to have a better understanding of costs for extended security guarantees. Full instrumentation of the C library aims to mitigate scenarios where the libc vulnerabilities can be exploited to corrupt the stack objects of the program or the library. We have experimented with both HMAC-SHA256 (using ISA acceleration) and SipHash-2-4 for integrity checks.

\subsection{Program-only Instrumentation}
In case only the program binaries are instrumented, both MAC implementations promise better execution times compared to unoptimised binaries (\codechunk{-O0}), where no register allocation takes place. As seen in Figure~\ref{fig:overheads}, only two benchmark programs with SipHash have produced slower execution times than the unoptimised versions. Considering a comparison between the basic register allocation without any instrumentation and our scheme compiled with the same optimisation level (\codechunk{-O2}), SHA256 backed by native ARMv8-A instructions has produced only \textit{13}\%, whereas SipHash yields \textit{23}\% overhead.

\subsection{With C Library Instrumentation}
We have observed higher performance costs for programs linked to an instrumented C library as expected. Compared to the naive scenario where both benchmark programs and libc are neither instrumented nor optimised, our implementation has still produced better execution times on average for the suite. Only three programs using HMAC-SHA256 and four programs with SipHash out of 14 benchmark executables have had slower execution times than non-optimised versions. In contrast to the basic register allocation bundled with \codechunk{-O2} optimisations, SHA256 and SipHash instrumentation have introduced \textit{33}\% and \textit{59}\% runtime overheads, respectively. Considering the binary sizes, instrumented C library with wrapper functions is only \textit{14}\% higher than the non-instrumented library file.\looseness=-1

\begin{figure}[tp]
	\centering
	\includegraphics[scale=0.21]{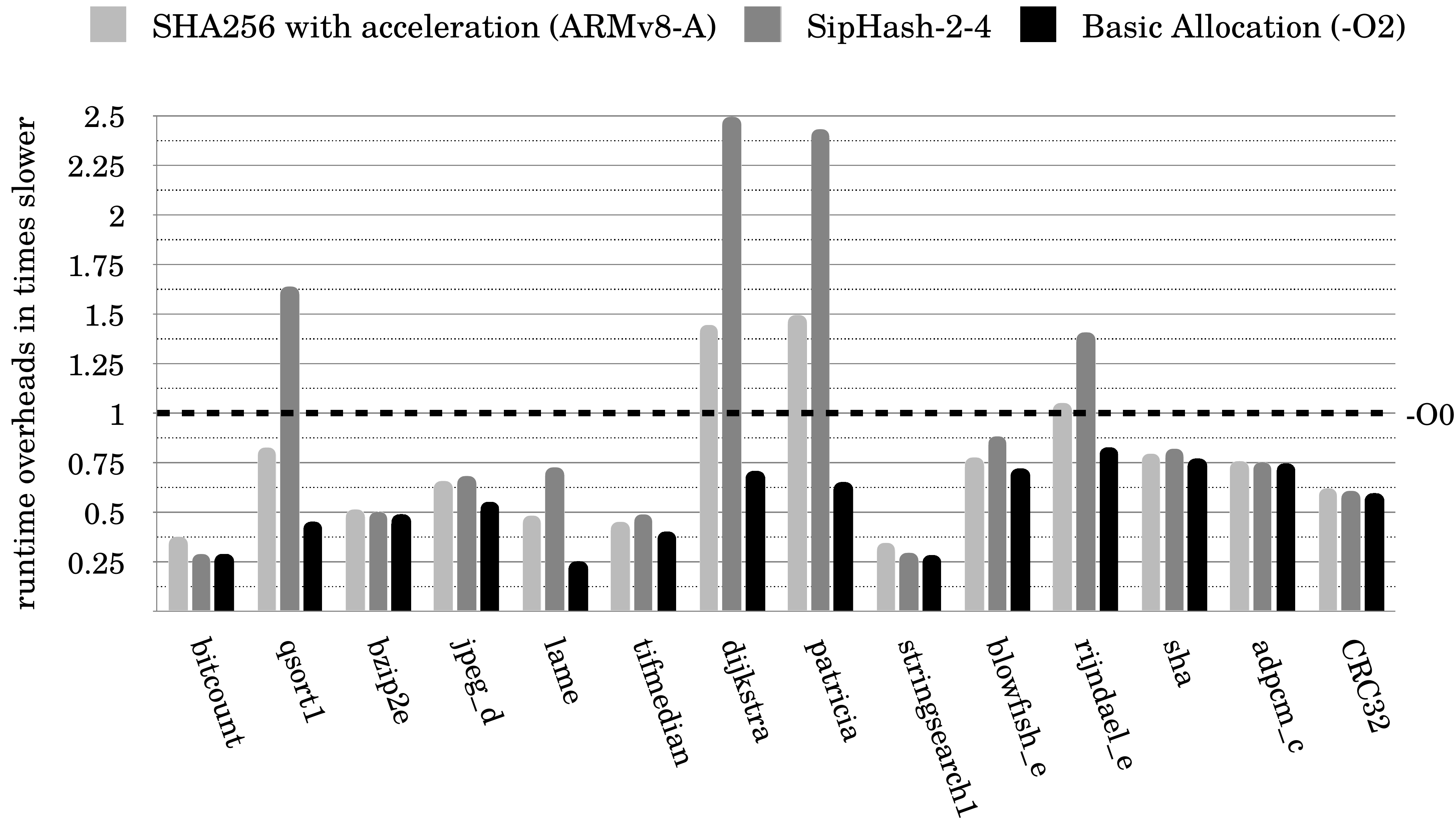}
	\caption{Runtime overheads with libc instrumentation.}
	\label{fig:overheads-libc}
\end{figure} 

Because the optimisation flags do not allow us to measure the performance impact of register allocations in isolation, we have used \codechunk{-O2} as the default optimisation level. Comparisons with basic register allocation create a baseline scenario to understand the standalone costs of additional integrity checks. On the other hand, experiments with unoptimised and non-instrumented programs aim to measure the compensation level by the register allocations of \scheme. We note that there are other types of optimisations included in the bundle contributing to the overhead compensation. For instance, inlining some functions not only avoids branching costs but also reduces tag calculations. This is due to the fact that the caller aggregates register operations of the inlined function. Overall, SipHash, with its reasonable overheads, proves to be a practical option for different CPU architectures without asking for any hardware change or acceleration. If available, similar to ARMv8.3-A, using native SHA instructions that provide around \textit{7x} speed-up would be a faster and more convenient option. Depending on the CPU features, both options can thus be practically used to ensure the integrity of register data on the stack since the overheads are within very small fractions of optimised times (\codechunk{-O2}) for most programs.

\section{Related Work}
This section reviews relevant work previously proposed and discusses how \scheme differs from them.

\subsection{Control-Flow Protections}
Many studies have been proposed to mitigate control-oriented attacks. Stack canaries~\cite{Cowan1998StackGuardOf} place random values next to return addresses to detect overflown buffers onto them. However, they capture neither attacks on forward-edge control objects (indirect branches) nor targeted modifications of return addresses (e.g.,~format-string attack). Control-flow integrity (CFI) techniques~\cite{Abadi2005,Niu2013ModularIntegrity} do not bother with how the corruption occurs by checking the validity of branch destinations using the control-flow graph (CFG). Although a shadow stack can assist for a fully precise backward-edge CFI, forward-edge targets can only be approximated depending on what is decidable and computable at compile-time. In contrast, our scheme proposes a more precise approach focusing on the integrity of forward-edge control objects rather than approximating their values. Similarly, code-pointer integrity (CPI)~\cite{Kuznetsov2014Code-pointerIntegrity} focuses on integrity assurance by placing code pointers on a safe stack, the location of which is hidden through randomisation within the same process memory. However, integrated attacks that reveal the location of the safe stack can simply circumvent its promises~\cite{Evans2015MissingIntegrity}. \scheme does not need to worry about those attack scenarios as it does not require isolation or hiding data within the same process memory.

\subsection{Mitigation of Data-Oriented Attacks}
Control-flow protections fail to address attack scenarios where the attacker does not necessarily touch any code pointers (i.e.,~DOP). Unfortunately, those data-oriented attacks continue to stay as the most challenging attack class without a practical protection deployed in the wild. Miguel~et~al. \cite{Miguel2006SecuringIntegrity} have proposed the first instance of data-flow integrity (DFI) schemes against those attacks. DFI checks whether any data object to be used at runtime is defined by an instruction known by compile-time flow-sensitive reaching definitions analysis. As a software-based approach, DFI suffers from excessive instrumentation of almost every memory access to protect both program and instrumentation data. A more coarse-grained technique with better performance in return for the loss of precision, write integrity testing (WIT)~\cite{Akritidis2008PreventingWIT} instruments only write instructions to prevent them from modifying objects that are not in the set of flow-insensitive points-to analysis. On the other hand, two relevant studies PointGuard~\cite{Cowan2003PointGuard:Vulnerabilities} and data space randomisation (DSR)~\cite{Bhatkar2008} mask data objects with random values and unmask them prior to their use. The main goal is to make corrupted values useless for an attacker that does not know masking values. Although masking memory representations harden the attacks leveraging pointers addresses, the attacker can still manipulate branch decisions made based on boolean or value range comparisons. Differently, \scheme detects the corruption of critical data objects under stronger adversary assumptions (e.g.,~memory disclosure), regardless of whether they are useful or not to the attacker.

\subsection{Hardware-Assisted Protections}
Regardless of their coverage, those protections must first ensure the integrity of their instrumentation data (e.g.,~shadow stack). But this is a challenging task without special hardware assistance. Hardware-assisted schemes can provide better performance and protection against both control~\cite{Davi2015,Christoulakis2016} and data~\cite{Song2016HDFI:Isolation,Nyman2017} attacks. However, those academic proposals are not usually adopted in practice as they require changes in CPU architectures, and the manufacturers do not implement them due to various reasons. Furthermore, already available features provided specific CPUs to protect instrumentation data, such as Intel MPX\cite{IntelCorporation2013IntroductionExtensions} and MPK~\cite{Corbet2015}, are shown to have high instrumentation or switch overheads despite their strong security promises~\cite{Burow2019SoK:Stacks}. In contrast, \scheme promises the same level of integrity assurance as an instrumentation-only solution using very basic primitives that are available in any CPU. This makes our scheme applicable to both legacy and modern architectures for a broad spectrum of devices, from high-end processors to low-end embedded systems.

\subsection{Cryptographic Protections}
MACs are first used by CCFI~\cite{Mashtizadeh2015CCFI:Integrity} to mitigate control-oriented attacks on x64 architectures. A CBC-MAC is computed and placed alongside each control object on the memory. To harden replay attacks, CCFI extends each 48-bit code pointer to a 128-bit AES block with additional information (e.g.,~frame address). The authors leverage Intel's AES-NI extensions to speed up MAC calculations. As a drawback, CCFI occupies most of FPRs (i.e.,~11 out of 16 XMM registers) for AES round keys. 
A similar work~\cite{Liljestrand2019PACAuthentication} presents the use of new pointer authentication (PAC) features provided by ARMv8.3-A. PAC tags are generated from effective address bits (39-bit) and squeezed into the (unused) upper part (24-bit) of the address word, which makes them susceptible to brute-force scenarios due to the short size. PAC associates return addresses with the stack pointer to avoid replay (pointer substitute) attacks. PAC does not provide any mechanism to detect corruption of a primitive variable, for instance, a condition variable overflown by an adjacent buffer. Similar to CCFI, PAC authenticates pointers in a standalone way with a separate MAC tag for each, in contrast to our work that digests many control and data objects using a single tag. Furthermore, both ideas are only applicable to specific CPU models.
Lastly, another recent scheme, ZipperStack~\cite{Li2020ZipperShadow} creates a chain of tag to protect return addresses on the memory. This study protects only return addresses and does not cover other control or data attacks targeting indirect branches or critical program variables. Similarly to PAC, ZipperStack stores MACs on the upper (24-bit) space of word, which provides weaker protection. Apart from their limited coverage, none of those schemes leverages the security and performance features of CPU registers as means for protecting critical objects in use.\looseness=-1

\section{Discussion}
\label{sec:discussion}
In this section, we present a discussion of certain design decisions of \scheme, including further extensions and future CPU design features that would complement our scheme.

\subsection{Chained vs Independent Frames}
Given that \scheme uses a keyed hash, it is not a strict requirement to include the previous tag in the tag of the next frame. In other words, we could have chosen to independently secure each call frame, rather than chaining them together. This section will briefly look at the reasons for and against this design desertion.

For a program with a call stack strictly following LIFO, we could have relied solely on a single (unkeyed) hash for the stack \textit{integrity} by chaining frames. This is because such a program can ensure that any CPU state restored from the stack complies with the hash register first. However, there are many legitimate cases where the register hosting the head of the chain has to be saved to/restored from program memory without our instrumentation, for example, setjmp/longjmp, exception handling and user-managed threading instances. They all oblige us to rely on the MAC key instead of a single hash.\looseness=-1

Despite its redundancy for integrity assurance, we have chosen the chained approach over independent frames to prevent \textit{replay attack} scenarios. With independent frames, the attacker can simply replay a call frame (and its aligned tag) for a different function call or context. However, with a chained approach, replaying for a different call context will not work since the tag register provides a very tight representation of the execution context, including all functions calls waiting to be returned. Even though setjmp/longjmp and user-managed thread instances might still provide a small window, it is very unlikely for the attacker to find a useful tag he can replay. This is because he needs a more coarse-grained stack-size image this time. Also, he will have fewer options; for example, he can exploit only setjmp/longjmp instances instead of function prologues/epilogues.

The only downside of a chained approach is occupying an additional register, which has to be excluded from allocations. This might be an issue for some legacy or primitive architectures that suffer from register scarcity. In such cases, the independent frame approach can be preferred to avoid the use of an extra (tag) register. To harden replay attacks without chained frames, we suggest including the stack pointer and a static function identifier or a nonce generated by the compiler as an immediate value in tag calculations. These two parameters provide a good approximation of the context by describing the current stack depth and returning function. The attacker cannot modify the function identifier, thanks to the code integrity. Also, the stack pointer would be safe by default on a register that can be saved to memory for the same reasons as the tag register.

\subsection{Primitive Devices and Register Scarcity}
\scheme uses security scores to distinguish critical variables and prioritise them for available registers under register pressure scenarios. However, it is difficult to observe such use cases with a modern CPU core providing a register file consisting of 48-64 (16/32 GPRs and 32 FPRs) registers with sizes up to 2kB. Hence, our selection process actually serves more primitive architectures suffering from register scarcity (e.g.,~6-8 GPRs with no FPRs). In such a case, our security scores aim to accommodate at least all critical objects in registers. But if there is still a critical object (e.g.,~condition variable) left in the memory, the compiler would display a warning; so the programmer can review the code. Despite being ignored by some compilers, the programmer can use the \codechunk{register} keyword in C to annotate which variables to protect. 
A different approach for CPUs with register scarcity can be adapting \scheme as a local register allocation scheme. Such a scheme would mitigate the register pressure problem by enabling the reuse of registers at a smaller (basic block) level in return for higher overheads. \looseness=-1

We have designed \scheme as an architecture-agnostic solution to make it applicable to a wide range of systems, even with the most resource-constrained devices in mind; for example, a 16/32-bit MCU with no security at all, but might be still prevalent in critical systems. By just relying on a flash program memory and a few GPRs, we can reduce the attack surface significantly with shorter keys and checksums against less strong adversaries.

\subsection{Future CPU Architectures}
Although \scheme is designed to fit existing CPU architectures, we would like to see CPU manufacturers incorporate some of these ideas into their designs in the future.

If the next generation of CPUs were to include the necessary registers and maybe even hardware acceleration of a suitable cryptographic MAC function, \scheme could be implemented at the hardware level through a single instruction. A bit vector-like operand can be used to describe which registers to include in the MAC, and the new instruction can then run all the necessary calculations within the CPU.

Furthermore, similar to Itanium (IA-64) architecture providing 128 GPRs and 128 FPRs, CPU manufacturers can consider expanding their register files as trusted storage and adopting \textit{register windows} to zero out the performance costs in return for space overhead within the CPU. Register windows, which are designed to avoid the cost of spilled registers on each call by making only a portion of registers visible to the program, can actually benefit our scheme more than its original purpose by eliminating cryptographic calculations. For example, with a window size of 32 (from 128 registers), \scheme would not incur any overheads for a program that has no call down deeper than four calls.

\subsection{Further Extensions}
\scheme covers attack scenarios that require modifying a stack object in the first instance. Due to the integrity assurance of stack pointers, most illegitimate accesses to other memory sections outside the stack would also be mitigated. However, there might still be some options for the attacker not addressed by the protection of stack pointers, such as overflowing a global or a heap array to corrupt another variable next to it. But thanks to the key register and MAC properties, we can extend our scheme to ensure the integrity of those objects. For example, we can allocate a tag address next to each global variable or composite data that will host a digest of them. We can update this tag at each legitimate (re)definition of those variables and verify when used.

\section{Conclusion}
This paper presents \scheme, a novel and practical scheme that leverages CPU registers to mitigate control- and data-oriented attacks targeting critical runtime objects, for instance, return addresses, function pointers and condition variables. Our protection relies on the immunity of registers from memory corruptions as unaddressable storage units. Despite their heavy use by compiler optimisations, CPU registers have not been previously used for security purposes due to their limited capacity for hosting variables program-wide or voided immunity across function calls thanks to values saved to the stack.\looseness=-1

\scheme addresses this challenge with a two-step proposal. First, during register allocations, it prioritises variables that are more likely to be targeted; so they stay safe while in use. Second, when those registers are saved to the stack because of a new function call, we compute a keyed hash to ensure they are restored without any corruption. Those integrity checks enable reusing the same register file as secure storage repeatedly for each function call, without having to occupy registers across function boundaries.

Although \scheme is designed as a software-based approach to be practical, it makes strong security promises using a very basic hardware primitive, CPU registers. This makes our scheme applicable to a very broad range of devices from high-end to low-end without asking for any special hardware features. Our experiments on ARM64 shown that register allocations can improve both the security and performance together with a surplus within the range of 13\% (with SHA extensions) to 23\% (SipHash) on average compared to purely performance-based optimisations.

\scheme is the first scheme that proposes the systematic use of CPU registers for security. It builds a practical protection with building blocks that are available in most computers, such as code integrity, registers and MAC calculations that can be expressed by any CPU ISA.

\bibliographystyle{plain}
\bibliography{references}

%%%%%%%%%%%%%%%%%%%%%%%%%%%%%%%%%%%%%%%%%%%%%%%%%%%%%%%%%%%%%%%%%%%%%%%%%%%%%%%%
\end{document}